# Gear Fault Diagnosis Based on Gaussian Correlation of Vibrations Signals and Wavelet Coefficients

Amir Hosein Zamanian[a], Abdolreza Ohadi[b*]

[a] MSc Student of Mechanical Engineering, zamanian.amir @aut.ac.ir
[b] Associate Professor of Mechanical Engineering, a_r_ohadi @aut.ac.ir
doi:10.1016/j.asoc.2011.06.020

Acoustics Research Lab., Mechanical Engineering Department, Amirkabir University of Technology (Tehran Polytechnic), Hafez Ave., 424, Tehran, Iran

* Corresponding author: Tel: +98 21 64543484, Fax: +98 21 66419736, email address: a_r_ohadi @aut.ac.ir

**Abstract**

The features of non-stationary multi-component signals are often difficult to be extracted for expert systems. In this paper, a new method for feature extraction that is based on maximization of local Gaussian correlation function of wavelet coefficients and signal is presented. The effect of empirical mode decomposition (EMD) to decompose multi-component signals to intrinsic mode functions (IMFs), before using of local Gaussian correlation are discussed. The experimental vibration signals from two gearbox systems are used to show the efficiency of the presented method. Linear Support Vector Machine (SVM) is utilized to classify feature sets extracted with the presented method. The obtained results show that the features extracted in this method have excellent ability to classify faults without any additional feature selection; it is also shown that EMD can improve or degrade features according to the utilized feature reduction method.

**1. Introduction**

Nowadays, vibration condition monitoring of industrial machines is used as a suitable tool for early detection of variety faults. Data acquisition, feature extraction, and classification are three general parts of any expert monitoring systems. One the most difficult and important procedure in fault diagnosis is feature extraction which is done by signal processing methods. There are various techniques in signal processing, which are usually categorized to time (e.g. [1, 2]), frequency (e.g. [3]), and time-frequency (e.g. [4, 5]) domain analyses. Among these, time-frequency analyses have attracted more attention because these methods provide an energy distribution of signal in time-frequency plane simultaneously, so frequency intensity of non-stationary signals can be analyzed in time domain.

Continuous wavelet transform (CWT), as a time-frequency representation of signal, provides an effective tool for vibration-based signal in fault detection. CWT provides a multi-resolution capability in analyzing the transitory features of non-stationary signals. Behind the advantages of CWT, there are some drawbacks; one of these is that CWT provides redundant data, so it makes feature extraction more complicated. Due to this data redundancy, data mining and feature reduction are extensively used, such as decision trees (DT) (e.g. [6]), principal component analysis (PCA) (e.g. [7]), independent component analysis (ICA) (e.g. [8-10]), genetic algorithm with support vector machines (GA-SVM) (e.g. [1, 2]), genetic algorithm with artificial neural networks (GA-ANN) (e.g. [1, 2]), Self Organizing Maps (SOM) (e.g. [11]), and etc.

Selection of wavelet bases is very important in order to indicate the maximal capability of features extraction for desired faults. As an alternative, Tse et al. [4] presented "exact wavelet analysis" for selection of the best wavelet family member and reduction of data redundancy. In this method, for any time frame, by defining cosine function between wavelet coefficients and the inspected signal, GA finds the scale parameter and some parameters related to wavelet family member (i.e. shape of wavelet). However, in the best conditions, GA finds only the most appropriate member of the wavelet family of the desired adaptive wavelets. This wavelet member may not be necessarily the best wavelet shape. Another approach is designing the wavelet based on similarity of wavelet and desired signal, but this method is not complete because it does not consider frequency properties of wave and signal. Lin and Zuo [12] suggested an adaptive wavelet filter based on Morlet wavelet. In their presented method, wavelet parameters are optimized based on the kurtosis maximization principle. Holm-Hansen et al. [13] presented a customized wavelet using the scaling function that has been derived from the actual impulse response of a ball bearing. Customized wavelet has shown better performance in comparison to other wavelets commonly used in detection of bearing faults.

Although exact wavelet analysis shows efficiency, but it has some drawbacks such as selection of scale in case of multi-component signals, equal weights of all samples participating in each time frame, treating of objective function with non-zero mean signals in each time frame and using of time consuming evolutionary algorithm which will be discussed in the next sections in more details.

Support Vector Machine (SVM) has been used in many applications of machine learning because of high accuracy and good generalization ability. The foundations of support vector machines (SVM) have been developed by Vapnik [14] in 1995 and are gaining popularity due to many attractive features, and promising empirical performance. The formulation embodies the structural risk minimization (SRM) principle, which has been shown to be superior to traditional empirical risk minimization (ERM) principle, employed by conventional neural networks [15]. Many researchers compared SVM with artificial neural networks (ANN) in fault diagnosis (e.g. [1, 2]) and generally showed that SVMs have better classification ability. Some researches (e.g. [6]) implemented different SVM algorithms to compare them in the field of gear fault diagnosis. In the current paper, linear SVM has been utilized to determine linear separability of extracted features.

In this paper a modified method for feature extraction based on exact wavelet analysis is presented to improve this analysis. A flowchart is shown in Fig. 1 for summarizing the proposed fault diagnosis procedure that has been used in this research.

The next sections of this paper are devoted to introduction of empirical mode decomposition, wavelet transform, exact wavelet analysis, modified exact wavelet analysis, support vector machine, two illustrative examples, experimental results and conclusions, respectively.

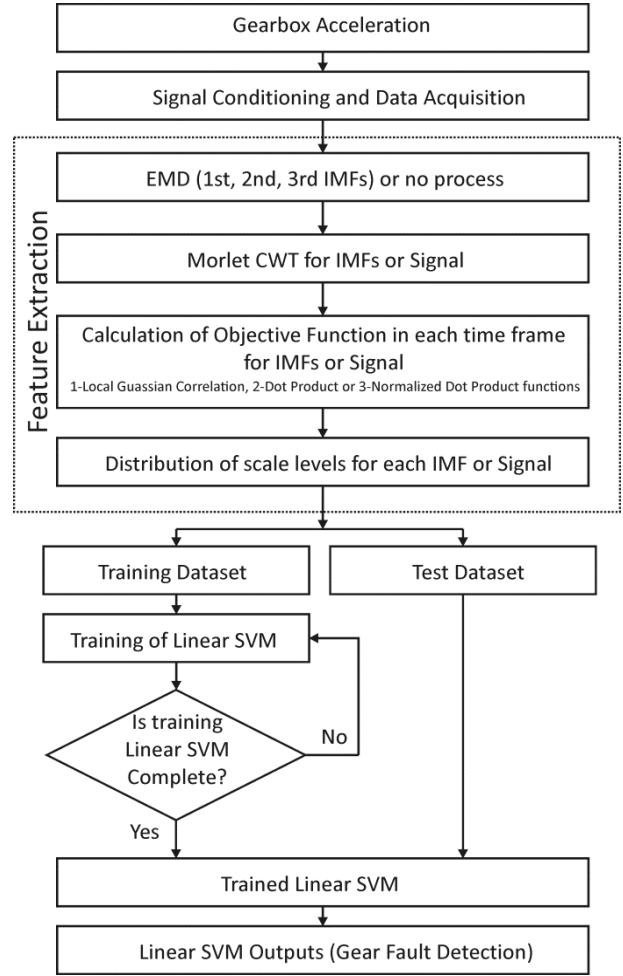

Fig 1. Flowchart of fault diagnosis system.

## 2. Empirical Mode Decomposition (EMD)

The empirical mode decomposition (EMD) presented by Haung et al. [16] is a decomposition method for nonlinear and non-stationary time-series into a finite set of oscillatory functions called the intrinsic mode functions (IMF). An IMF represents a simple oscillation with the following definition: (1) the number of extrema and the number of zero-crossings must either equal or differ at most by one, and (2) at any point, the mean values of the envelopes defined by the local maxima and the local minima are zero. Therefore, an IMF represents a simple oscillatory mode that has a variable amplitude and frequency as function of time. The readers are referred to [17] to find more details on EMD. The procedure to decompose a signal $x(t)$ into its IMFs is as follows:

1) Identify all the local maxima and connect them by a cubic spline to produce the upper envelope, $u(t)$.

2) Repeat the procedure for the local minima to produce the lower envelope, $l(t)$.

3) Calculate mean of two envelopes as $m_1(t) = (u(t) + l(t))/2$.

4) Extract $h_1(t)$ by subtraction of $m_1(t)$ from signal $x(t)$ (i.e. $h_1(t) = x(t) - m_1(t)$).

Ideally, $h_1(t)$ should satisfy the definition of an IMF, however, even if the fitting is perfect, a gentle hump on a slope can be amplified to become a local extremum in changing the local zero from a rectangular to a curvilinear coordinate system [17]. To eliminate riding waves and to make the wave profiles more symmetric, a process called *sifting* by Huang [17] should be followed. Therefore, the next step is:

5) Repeat the procedures 1 to 4 (sifting process), by replacement of $h_{1k}(t)$ instead of $x(t)$ (i.e. $h_{1k} = h_{1(k-1)} - m_{1k}$, $k = 1, 2, ...$). Here $h_{10}(t) = h_1(t)$ and $m_{1k}$ donates to mean of upper and lower envelopes of $h_{1(k-1)}$ in the $k^{th}$ iteration, and the process will continue until one of the stoppage criterion is met. These criterions are:

   a) A threshold of normalized squared difference in two successive iterations as defined below. If this value (i.e. $SD_k$) is smaller than a predetermined value, the sifting process will be stopped.

$$SD_k = \frac{\sum_{t=0}^{T} |h_{1k-1}(t) - h_{1k}(t)|^2}{\sum_{t=0}^{T} h_{1k-1}^2}$$

   b) The agreement of the number of zero-crossings and extrema. If the numbers of zero crossing and extrema are not equal or not differ at most by one, then $h_{1k}$ cannot be considered as an IMF, and EMD is finished.

6) The final $h_{1k}$ is the 1$^{st}$ IMF ( $p_1(t) = h_{1k}(t)$ )

By subtracting 1$^{st}$ IMF, $p_1(t)$, from signal $x(t)$, residual $r_1(t) = x(t) - p_1(t)$ is obtained. If the residual is taken as the new signal by iterating the above process, 2$^{nd}$, 3$^{rd}$... to n$^{th}$ IMFs can be calculated consequently. Thus, a decomposition of the data will result $n$ IMF and also the final residual $r_n(t)$.

$$x(t) = \sum_{i=1}^{n} p_i(t) + r_n(t)$$

## 3. Wavelet Transform

Wavelet transform has provided a power tool for signal analysis of transients. The continuous wavelet transform (CWT) of a signal $x(t)$ is defined as,

$$W_\psi(a,b) = \int_{-\infty}^{+\infty} x(t) \overline{\Psi}_{a,b}(t) dt \quad (1)$$

where

$$\Psi_{a,b}(t) = |a|^{-\frac{1}{2}} \Psi\big((t-b)/a\big) \quad (2)$$

is called daughter wavelet derived from mother wavelet $\Psi(t)$ and the bar denotes its complex conjugation and $a$ and $b$ are real-value parameters, called scale and translation, respectively. In this paper, Morlet wavelet [18], as a non-adaptive wavelet has been selected (Eq. (3)), which is considered as a common wavelet used in fault diagnosis.

$$\psi(t) = \frac{1}{\sqrt{2\pi}} \exp(-t^2) \cos(2\pi v_0 t) \qquad (3)$$

where $2\pi v_0$ equals 5 in this work.

## 4. Exact Wavelet Analysis [4]

Exact wavelet analysis, presented by Tse et al. [4], is aimed to provide a direct measure of the similarity in shapes between the daughter wavelet and the inspected signal. The "normalized dot product" (cosine function) of the CWT and the inspected signal is adopted for measuring their similarity in shape, as shown in Eq. (4).

$$\cos(\mathbf{X}, \mathbf{C}) = \frac{\sum_{i=1}^{2n+1} x_i c_i}{\sqrt{\sum_{i=1}^{2n+1} x_i^2} \sqrt{\sum_{i=1}^{2n+1} c_i^2}}, \quad n = 1, 2, 3, \ldots \qquad (4)$$

In this equation, $\mathbf{X}$ and $\mathbf{C}$ represent vectors of signal and its wavelet coefficients in a definite scale. The variables $x_i$ and $c_i$ represent the elements of the vectors and $2n+1$ is the number of data. The calculated index from the fitness function provides a measure to evaluate the similarity of the two vectors not only in their magnitudes but also in their geometrical shapes. The higher the indexes of the fitness function, the more similar are the derived wavelet and the portion of the inspected signal. The index of the cosine function approaches to 1 indicating a perfect match, whilst the index approaches to zero shows a mismatch. Genetic algorithms are employed to optimize Eq. (4) as a function of scale and parameters related to wavelet shape that are used to generate a series of daughter wavelets. Therefore, the derived exact wavelets are able to match properties of the inspected signal well. In this method, in each time frame, only one scale among possible scales, which has the best fitness function, is selected. The final extracted features are based on distribution of selected scales obtained by GA.

## 5. Modified Exact Wavelet Analysis

As explained in section 4, in exact wavelet analysis, one scale is chosen in each time frame based on maximization of normalized dot product of the wavelet coefficients and the inspected signal. This selection does not guarantee that fault effects are considered in the selected features in case of multi-component signals. In this paper it has been tried to solve this deficiency. Some drawbacks on exact wavelet analysis that are investigated in this paper are as follows:

1) In the exact wavelet analysis, definition of normalized dot product of the wavelet coefficients and the inspected signal in a time window is based on the equal weight of samples. Indeed, the center sample in window has equal effect with the samples at the boundaries of the window. This is not necessarily the best choice.

2) In practical multi-component signals, exact wavelet selects only one scale corresponding to one component of the signal. This component may not necessarily be due to the faults of the system.

3) In any time frame, the normalized dot product represents a good criterion for finding similarity of zero mean signals and its wavelet coefficients. However, for non-zero mean signals, this measure cannot be as effective as zero mean signals.

4) The use of GA for each time frame is a time-consuming process.

In this paper, some solutions are suggested as a new method for feature extraction based on maximization of "*local Gaussian correlation*" function of wavelet coefficients and signal. These solutions are as follows:

1) To solve the problem of equal weights, Gaussian weights is presented. The effect of samples, from center of time frame to the ends, reduces according to a Gaussian function from one to zero.

2) In practical applications, especially for vibration signals, there is a probability that the selected scale has not been caused by any faults and effects of the faults may be neglected. Indeed, some of the signal components are due to unbalanced mass, misalignment, looseness, bearing and gear faults, resonance, etc. that can affect the analysis. In order to solve this problem, usage of empirical mode decomposition (EMD) of signal is suggested. By using EMD, primarily, each vibration signal is decomposed to intrinsic mode functions (IMFs) that are almost mono-component and orthogonal. Then the procedure of exact wavelet analysis is implemented for each IMF. In this way, the probability of omitting scales which are due to the faults is eliminated or at least is reduced.

3) The "*normalized dot product*" function can be replaced with "*local correlation*" or "*dot product*" functions. In order to use these functions for non-zero mean time frames appropriately, subtraction of mean of samples in the selected time frame, and its corresponding wavelet coefficients can provide a solution. The suggested function is not bounded and the greater value means more similarity.

In the modified method, for each selected time frame with length of $2n+1$ samples, the similarity of wavelet coefficients *W(a,b)* and corresponding signal *x(t)* is calculated by local correlation function, which is defined in Eq. (5).

$$f(\mathbf{X}, \mathbf{C}) = \sum_{i=1}^{2n+1} w_i (x_i - \bar{x})(c_i - \bar{c}) \tag{5}$$

In this equation, $\mathbf{X}$ and $\mathbf{C}$ represent vectors of signal and its wavelet coefficients in a definite scale, in a time frame, respectively. Also, $\bar{x}$ and $\bar{c}$ are mean of signal and wavelet coefficient in selected time frame and $w_i$ is contributing weight of each sample in time frame. In the case of local correlation or dot product all of contributing weights equals one (i.e. $w_i = 1$) and in the case of Gaussian weights, $w_i$ can be calculated from Eq. (6).

$$w_i = \exp\left(-(i-(n+1))^2 / \tau^2\right) \tag{6}$$

Parameter $\tau$ in Eq. (6) can be determined by considering one percent participation of latest samples in the time window as shown in Eq. (7).

$$\tau = \frac{n}{\sqrt{2\ln 10}} \approx \frac{n}{2.146} \tag{7}$$

4) The GA optimization can be neglected in price of neglecting any adaptive wavelet. This can be done by using an appropriate non-adaptive wavelet and optimization of scale discretely with desirable resolution. So in each time frame, only one parameter (i.e. scale) should be optimized. The advantage of this method is that the optimization algorithm can be omitted with comparing objective function (i.e. local Gaussian correlation or normalized dot product function) in a finite number of discrete scales. Therefore, the computation effort has been significantly decreased. The calculated index in Eq. (5) provides a boundless measure to evaluate the similarity of the two unnecessarily zero mean vectors not only in their magnitudes but also in their geometrical shapes. The higher index of the function, the more correlation is derived. Let's clarify the difference of two functions in Eq. (4) and Eq. (5). It is clear that for constant signal in a time frame for example, Eq. (5) equal zero whereas Eq. (4) is not necessarily zero. In Eq. (4) the greater value of the constant signals leads the greater value of the function whereas in Eq. (5) the magnitude of constant signal doesn't affect the measure of similarity and the value remains zero. So Eq. (5) doesn't consider the effect of constancy and it focuses on similarity of the shape of signal and wavelet coefficients and higher wavelet coefficients, simultaneously. In case of local Gaussian correlation the effects of center of participating signal is more than phenomena that occurred before or will occur in futures.

## 5. Support Vector Machine

The concept of support vector machine is extensive. A brief introduction of SVM is presented here and the readers are referred to references [15] and [19] for more details. Without loss of generality, the classification problem can be restricted to consideration of the two-class problem. SVM can be considered to create a line or hyper-plane between two sets of data for classification. Consider the problem of separating the set of training vectors belonging to two separate classes:

$$D = \{(\mathbf{x}^1, y^1),...,(\mathbf{x}^l, y^l)\}, \quad x \in \mathbf{R}^n, y \in \{-1,1\}, \tag{7}$$

with a hyper-plane,

$$\mathbf{w}.\mathbf{x} + b = 0. \tag{8}$$

In the case of two-dimensional situation, the action of the SVM can be explained easily. In this situation, SVM find a line which separates two classes of data (feature sets). This line separates data into two parts so that the data on the right hand belong to one class (Class A) and the data on the left hand belong to the other class (Class B). Many lines may have the ability to separate data completely. However, SVM find the line that has the maximum Euclidean distance between the nearest data to this line both in Classes A and B. The data, which has the minimum distance to this line, are called support vectors (SVs) that are shown in Fig. 2. Since training SVM with SVs in case of linear separable data is sufficed, the rest of data can be neglected.

The SVs are located in two parallel lines, which are parallel to the separating line. The margin equations for class A and B are:

$$\mathbf{w}.\mathbf{x} + b = 1 \quad \text{(Class A)}, \tag{9}$$

and

$$\mathbf{w}.\mathbf{x} + b = -1 \quad \text{(Class B)}. \tag{10}$$

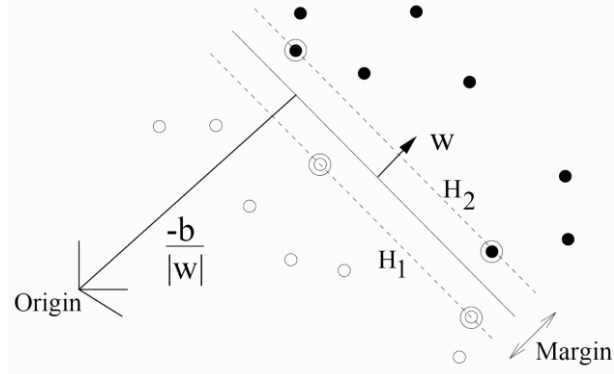

Figure 2. Linear separating plane for classification, the support vectors are circled [19].

Once the SVM has been trained, the decision function in Eq. (11) determines the class of each test sample and indicates its position related to the decision boundary.

$$f(x) = \text{sgn}(\mathbf{w}.\mathbf{x} + b), \tag{11}$$

The SVM training is obtained by optimizing an objective function that is presented in Eq. (12).

$$L = \frac{1}{2}\|\mathbf{w}\|^2 - \sum_{i=1}^{l} \alpha_i y_i (\mathbf{w}.\mathbf{x_i} + b) + \sum_{i=1}^{l} \alpha_i, \tag{12}$$

where $l$ is the number of training sets, and $\alpha_i$ is Lagrange multipliers' coefficients obtained by the following constraints.

$$y_i(\mathbf{w}.\mathbf{x}_i + b) \geq 1, \tag{13}$$

The solution can be obtained as follows:

$$\mathbf{w} = \sum_{i=1}^{l} \alpha_i y_i \mathbf{x}_i = \sum_{i=1}^{l} v_i \mathbf{x}_i, \tag{14}$$

where

$$v_i = \alpha_i y_i \tag{15}$$

Substitution of Eq. (14) into Eq. (11) leads to Eq. (16).

$$f(x) = \text{sgn}\left(\sum_{i=1}^{l} v_i (\mathbf{x}_i.\mathbf{x}) + b\right) \tag{16}$$

The set of vectors is said to be optimally separated by the hyper plane if it has been separated without error and the distance between the closest vector to the hyper plane is maximal [15]. In the case of non-separable data with linear hyper plane, a hyper plane should be defined so that it allows linear separation in the higher dimension (corresponding to nonlinear separating hyper planes). The intent of this paper is to show that the extracted features are linearly separable by the presented method, so Linear SVMs are expected to have excellent classification performance and nonlinear SVMs can be neglected.

## 6. Two Illustrative Examples

In this section, in order to clarify the presented method, two examples for two synthetic signals are presented.

### 6.1. *Example 1*

Two synthetic harmonic signals $y_1(t)$ and $y_2(t)$ have been superimposed to each other and create $y(t)$ in an interval of $0 < t < 1$ with sampling rate of 1 kHz as shown in Eq. (17).

$$y_1(t) = \sin(40\pi t), \quad y_2(t) = 2\cos(80\pi t), \quad y(t) = y_1(t) + y_2(t). \tag{17}$$

The coefficients of CWT of signal $y(t)$ are shown in time-scale plane (Fig. 3(a)). It is clear that wavelet coefficients near scale 40 are related to $y_1(t)$ and near scale 20 are related to $y_2(t)$. In Fig. 3(b) the local Gaussian correlation of signal with wavelet coefficient has been shown in time-scale plane and the maximum value of correlation is shown with white points which have been appeared line-wise. The synthetic signal contains two components but the extracted scales, according to Gaussian-correlation window, indicate only one of the components of signal. All features extracted in Fig. 3(b) are related to the signal with the higher frequency (i.e. $y_2(t)$). It is one of the deficiencies of exact wavelet analysis when the signal is multi-component. In practical applications (i.e. gearbox vibration signals), there is this probability that the signal contains frequencies near to gear mesh frequency (GMF) or higher. In this condition, the best fitness of local Gaussian correlation or any other function may represent the features that are not caused by faults in gear. So, the EMD method has been done for signal $y(t)$ and the first two IMFs have been computed. Then, for these IMFs, CWT and local Gaussian correlation have been calculated (the same as done for $y(t)$) as shown in Fig. 4 and 5. In this condition, in a time frame, one scale has been selected for each IMF. Since each IMF contains a component of the signal, this decomposition provides the opportunity for all components to be selected, and the final feature can be extracted by mixing these results.

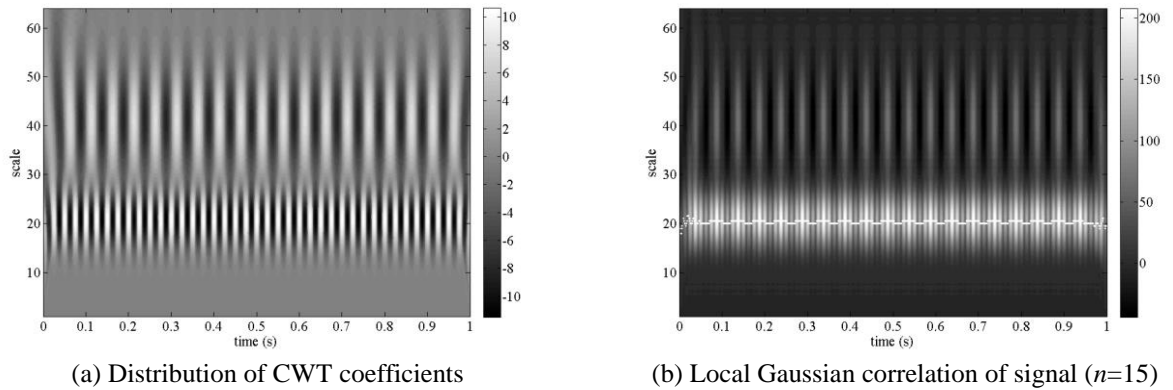

(a) Distribution of CWT coefficients    (b) Local Gaussian correlation of signal (*n*=15)

Figure 3. Energy distribution of signal $y(t)$.

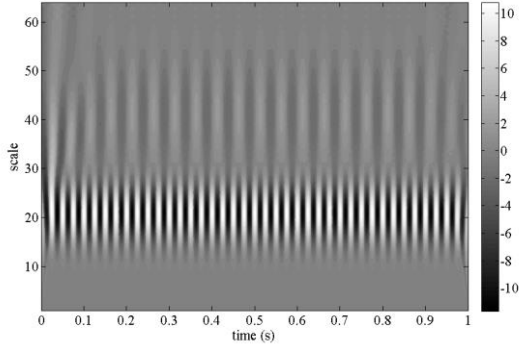 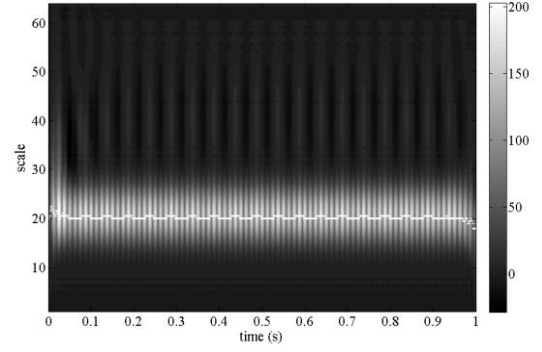

(a) Distribution of CWT coefficients　　　　　　(b) Local Gaussian correlation of signal (*n*=15)

Figure 4. Energy distribution of the first IMF.

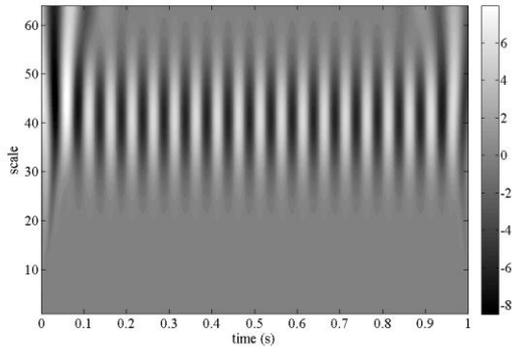 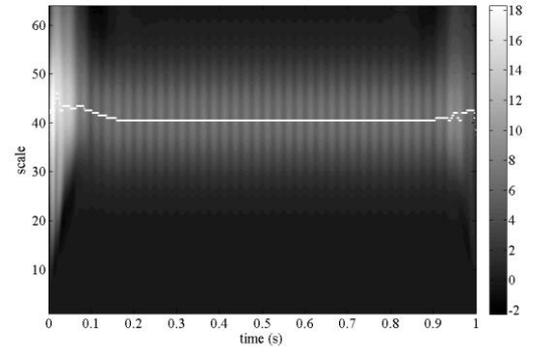

(a) Distribution of CWT coefficients　　　　　　(b) Local Gaussian correlation of signal (n=15)

Figure 5. Energy distribution of the second IMF.

### 6.2. *Example2*

The second example is devoted to a summation of a harmonic signal ($y_1(t)$) and a chirp ($y_2(t)$) in an interval of $0 < t < 1$ with sampling rate of 1 kHz (Eq. 18).

$$y_1(t) = \sin(40\pi t), \quad y_2(t) = \cos(2(10+80t)\pi t), \quad y(t) = y_1(t) + y_2(t). \tag{18}$$

In this example a chirp signal is intentionally chosen because in practical situations with controlled speed, even a small fault in gear can change rotating speed that containing chirp like characteristics due to the impact and inappropriate gear meshing. This is exaggerated in this example. Again, the coefficient of CWT and local Gaussian correlation for synthesized signal and its first two IMFs are shown in Figs. 6-8, respectively. It is obvious in Fig. 6(b) that local Gaussian correlation of signal, at least in the first half of chirp, cannot extract features of chirp component. Fig. 7(a) and Fig. 8(a) show that when frequency of chirp and harmonic signal at the end of the duration come close together, the EMD fails to decompose the signal appropriately. However, even in these conditions, local Gaussian correlation can track scales related to chirp and harmonic signal in a gentle manner.

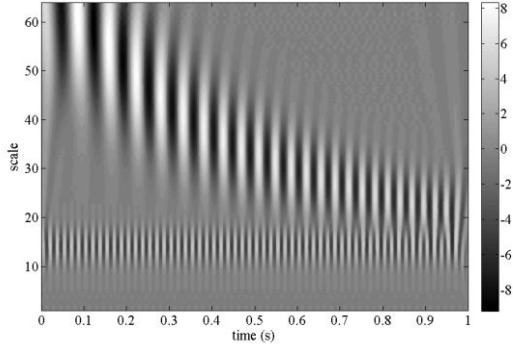 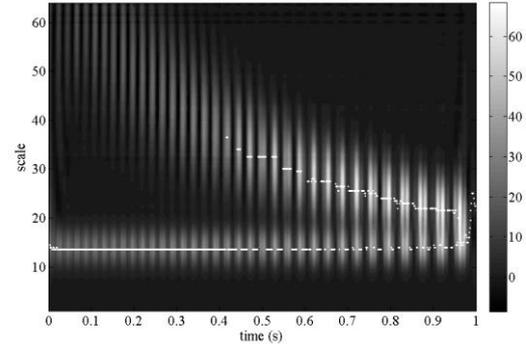

(a) Distribution of CWT coefficients    (b) Local Gaussian correlation of signal *(n=15)*

Figure 6. Energy distribution of signal $y(t)$.

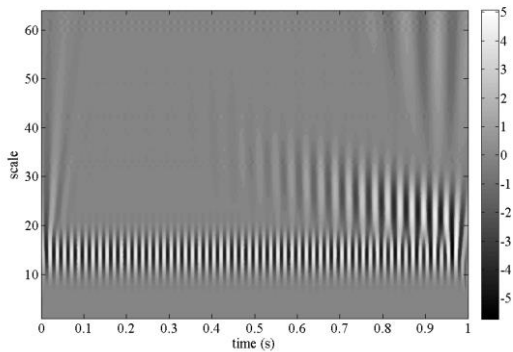 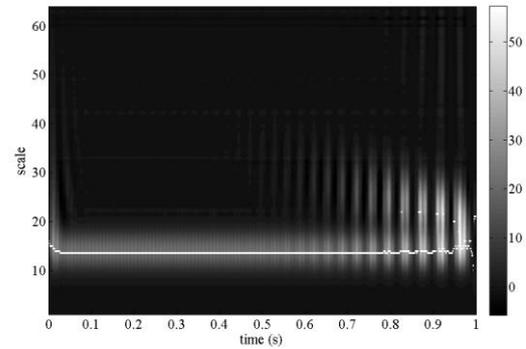

(a) Distribution of CWT coefficients    (b) Local Gaussian correlation of signal ($n$=15)

Figure 7. Energy distribution of the first IMF.

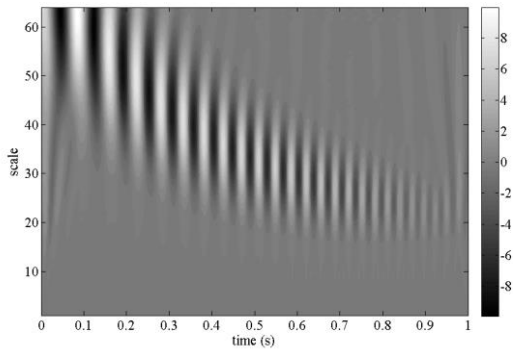 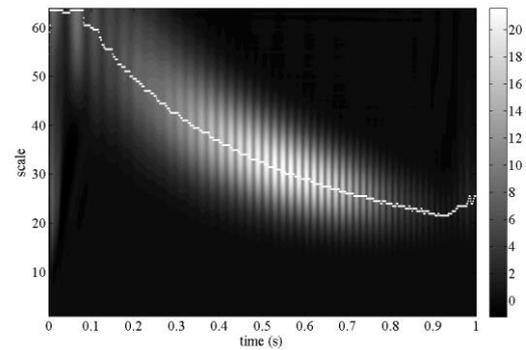

(a) Distribution of CWT coefficients    (b) Local Gaussian correlation of signal ($n$=15)

Figure 8. Energy distribution of the second IMF.

These two examples show that the greater coefficient of CWT cannot represent the best match of scale with a signal alone. The other fitness functions such as normalized dot product, Gaussian dot product, correlation coefficient have been checked. Among these functions, local Gaussian correlation represented better match than the other functions to follow signal in time-scale plane.

## 7. Experimental Results

To evaluate the efficiency of the presented method in fault diagnosis of practical applications, the vibration signals, which have been acquired from two gearboxes, are used in this paper.

## 7.1. Amirkabir Experimental Setup

Fig. 9 indicates a gearbox test setup designed in Amirkabir University of Technology (Tehran Polytechnic). The vibrating signals were obtained from 2D accelerometer (Analog Devices Inc.-ADXL210JQC) mounted on gearbox frame with sampling frequency rate of 10 kHz. The frequency content of measured acceleration is in the range of 0~5 kHz. These signals were fed to A/D converter (Advantech™ PCI-1710, 12-bit, 100kS/s) and were recorded by real-time workshop of MATLAB. The gearbox was driven by a 3-phase electromotor with nominal speed of 1420 RPM. The pinion ($N_1 = 15$) and wheel ($N_2 = 110$) teeth provide speed ratio of 7.33:1 for gearbox. This configuration provides GMF of $15 \times 1420/60 = 355\,\text{Hz}$. A disk brake system has been considered to provide appropriate load on the system. The gearbox was tested for three conditions; normal, chipped, and worn teeth as shown in Fig.10. In the chipped case, 50% area from top land of a pinion tooth profile to pitch surface has been eliminated with linear decreasing slop. In the worn case, material of the face and flank of three consequent teeth has been removed with thickness of $0.5\,\text{mm}$. It should be mentioned that the effect of the chipped and worn teeth gears are similar in signal and the difference is minor and these faults intentionally have been chosen to measure power discrimination of the presented methods. For each test condition, data was stored in duration of 10 sec and each acquired signal was divided to 80 segments with the length of 1250 samples. Each segment was used to extract feature sets, so in total 240 ($80 \times 3$) feature sets were obtained, 90 sets were used for training the SVM, and the remaining 150 feature sets were used for testing the performance of trained SVM.

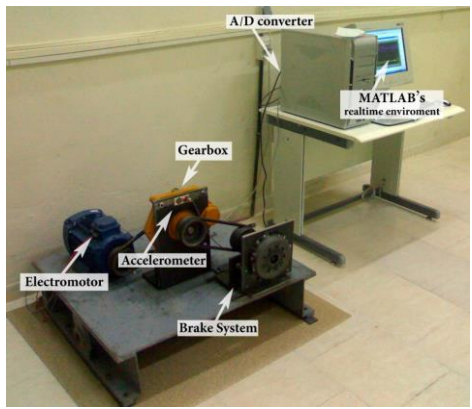
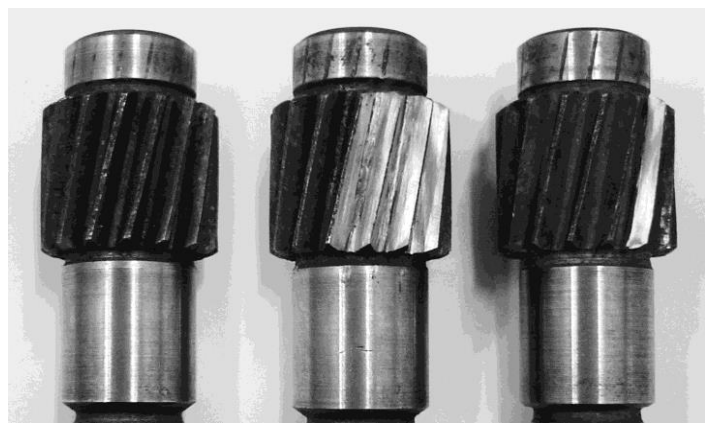

(a) Healthy  (b) Wear  (c) Chipped

Figure 9. The experimental setup test
(Acoustic Research Lab.,
Amirkabir University of Technology)

Figure 10. Gears conditions

For the first feature set, the coefficient of CWT and scales that maximize local Gaussian correlation function are shown in Fig. 11. The differences between gear conditions, obtained by local Gaussian correlation function, are obvious. For investigating the effects of EMD method, selected scales of the first three IMFs for the first segment are considered and plotted in Fig. 12, as a sample. The distributions of scale levels (suggested by Tse et al. [4]) in normal, chipped, and worn teeth gears are calculated, and provide $3 \times 32$ initial features for each segment. All of initial features obtained from IMFs are mixed together by summation of equivalent scale levels divided by the number of IMFs. This kind of summation sometimes may degrade features' value in classification because the features derived from each IMF have different dependencies to faults and need to be investigated further. So, finally with and without EMD process, 32 features are extracted as final feature sets. The mean of distribution of scale levels for all signal segments with and without EMD is plotted in Fig. 13.

The final training feature sets are used for training Linear SVM. The performance has been evaluated by test feature sets and has been tested with trained feature sets to evaluate the training success. In order to provide multiclass classification by SVM, one-versus-all approach has been used. The results are represented in Tables 1-3. For all gear conditions (healthy, chipped, and worn) in Table 1, local Gaussian correlation has the best match and dot product is on the second. 5.33% improvement has been achieved by local Gaussian correlation in comparison to normalized dot product. The classification performance by using EMD has been increased or decreased depend on the cases. Using EMD has reduced the classification performance in worn teeth condition by about 0.7% with first three IMFs and 1.3% with first four IMFs and 0.7% in chipped tooth with first four IMFs. The results may seem contrary to the expectation of EMD at first, and can be justify by this reason that, all IMFs of signal are not related to fault so summation of features of non-related components (i.e. related IMFs) with related components (i.e. non-related IMFs) in order to reduce features, degrade power of discrimination of the final extracted features. In all of the cases in Table 1, local Gaussian correlation is dominated to the others.

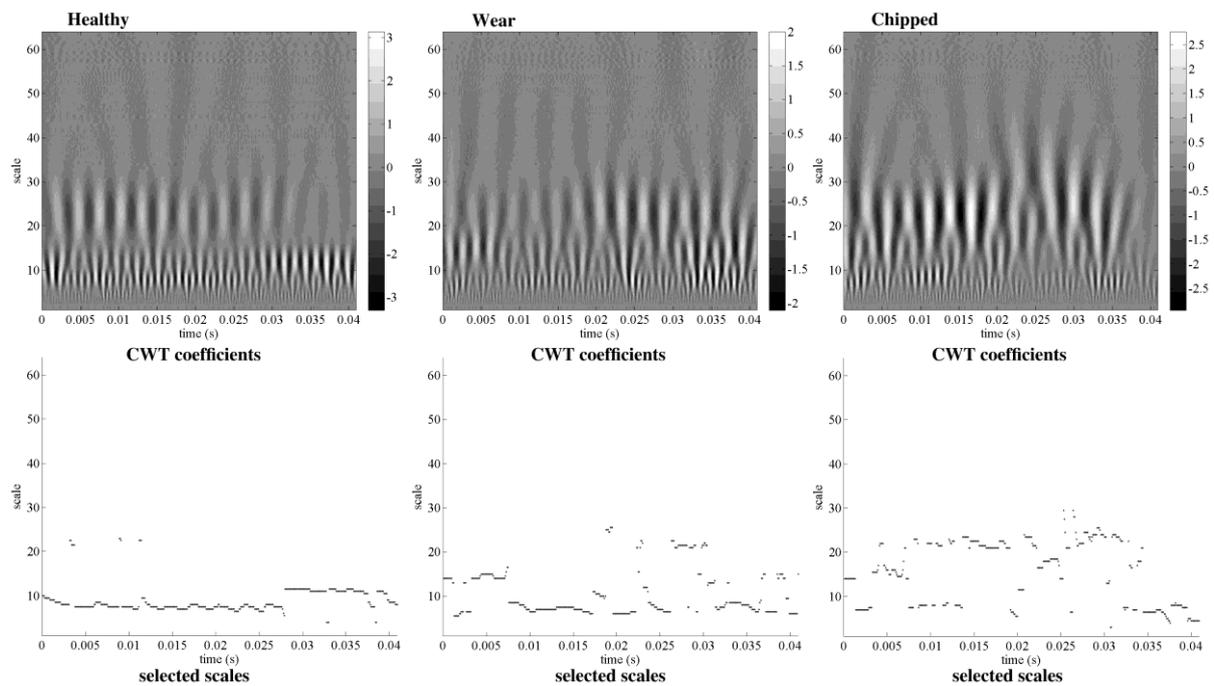

Figure 11. Distribution of CWT coefficients and selected scales for the first signal segments.

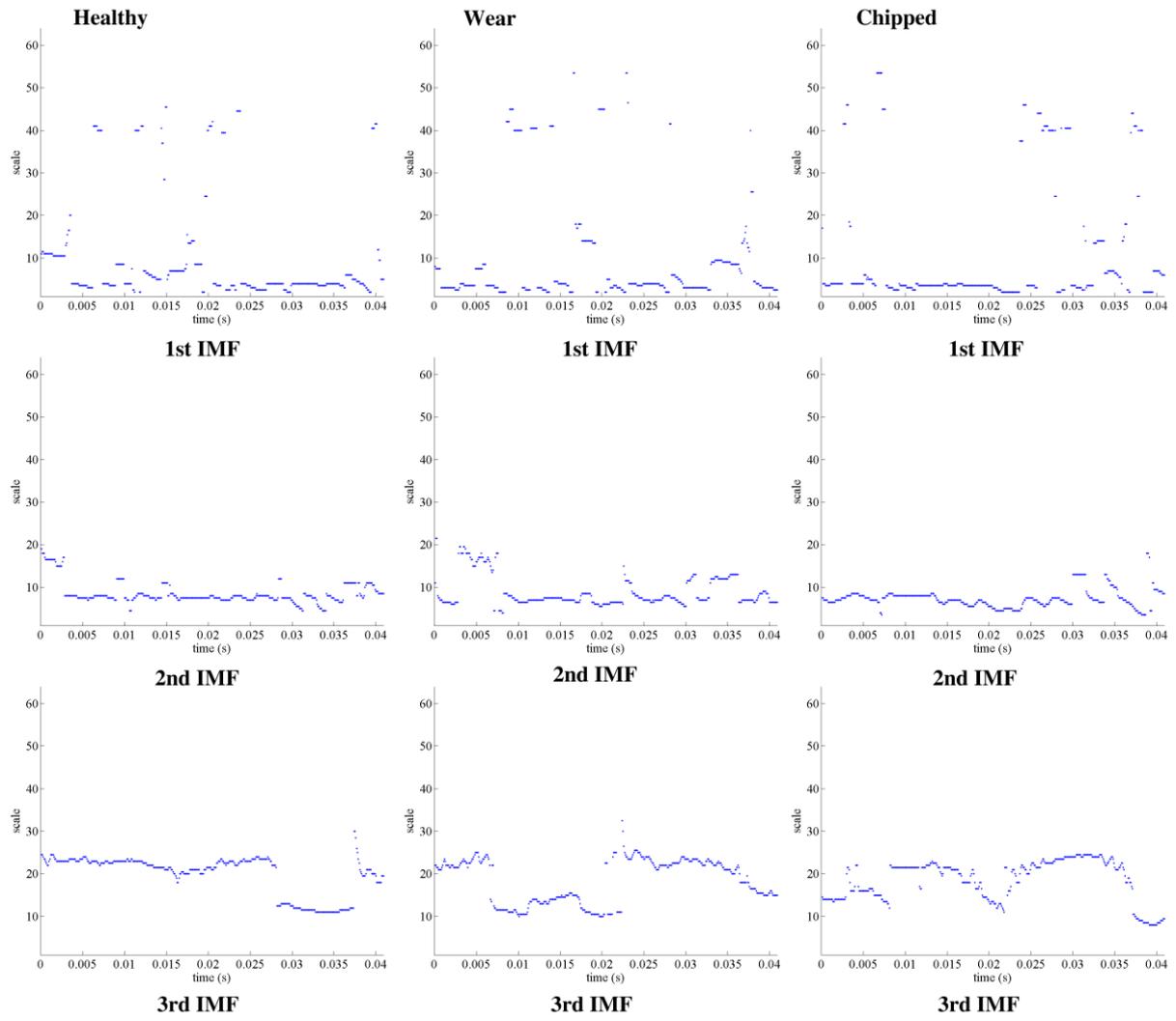

Figure 12. Selected scales by local Gaussian correlation for the first signal segments.

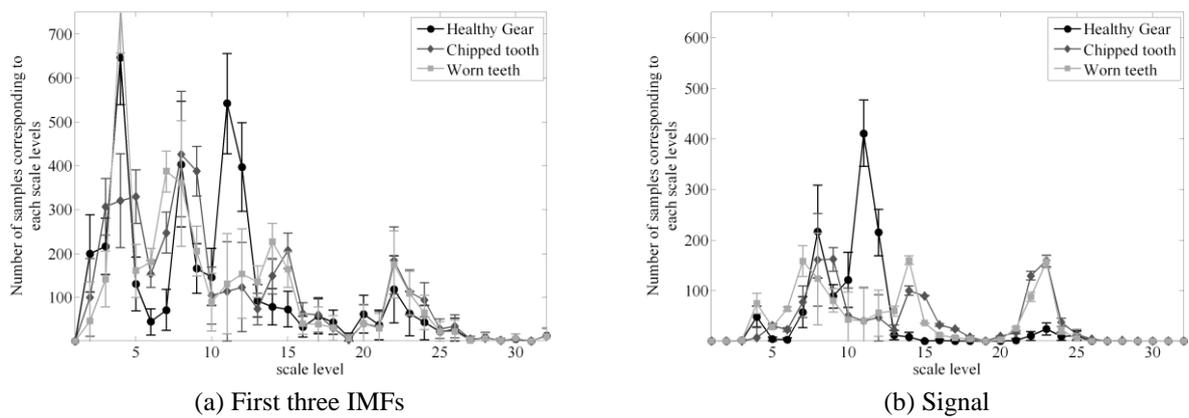

(a) First three IMFs  (b) Signal

Figure 13. Mean of distribution of scale levels of all signal segments.

Table 1. Classification performance of Linear-SVM (signal).

| Objective Function | Healthy | | Chipped tooth | | Worn teeth | |
|---|---|---|---|---|---|---|
| | Train success (%) | Test success (%) | Train success (%) | Test success (%) | Train success (%) | Test success (%) |
| Local Gaussian correlation | 100 | 100 | 100 | 99.33 | 100 | 100 |
| Dot product | 100 | 100 | 100 | 98.00 | 100 | 97.33 |
| Normalized dot product | 100 | 100 | 100 | 98.00 | 100 | 94.66 |

Table 2. Classification performance of Linear-SVM with first three IMFs.

| Objective Function | Healthy | | Chipped tooth | | Worn teeth | |
|---|---|---|---|---|---|---|
| | Train success (%) | Test success (%) | Train success (%) | Test success (%) | Train success (%) | Test success (%) |
| Local Gaussian Correlation | 100 | 100 | 100 | 99.33 | 100 | 99.33 |
| Dot product | 100 | 100 | 100 | 98.66 | 100 | 99.33 |
| Normalized dot product | 100 | 100 | 100 | 92.00 | 100 | 93.33 |

Table 3. Classification performance of Linear-SVM with first four IMFs.

| Objective Function | Healthy | | Chipped tooth | | Worn teeth | |
|---|---|---|---|---|---|---|
| | Train success (%) | Test success (%) | Train success (%) | Test success (%) | Train success (%) | Test success (%) |
| Local Gaussian Correlation | 100 | 100 | 100 | 98.66 | 100 | 98.66 |
| Dot product | 100 | 100 | 100 | 99.33 | 100 | 98.00 |
| Normalized dot product | 100 | 100 | 100 | 95.33 | 100 | 94.00 |

### 7.2. Lemmer's Data

In this case study, the machine under investigation is a pump that is driven by an electromotor. The speed of incoming shaft is reduced by two decelerating gear-combinations. Measurements were repeated for two identical machines; the first pump had severe gear damage (pitting), whereas the second pump was fault free [20]. Seven uni-directional accelerometers were used to measure the vibration near different structural elements of the machine (shaft, gears and bearings) [20]. The pinion teeth number was 13 and the driving shaft speed was 997 RPM, which results in a GMF of $13 \times 997/60 = 216 \, \text{Hz}$. The first three channels correspond to three different measurement directions at the position near the incoming axis. Sensors 4 to 7 were mounted on the gearbox casing. Sensors 4 and 5 were mounted near the first set of deceleration gears (sensor 4 on the upper part and sensor 5 on the lower part of the casing). Sensors 6 and 7 were mounted near the second set of gears (sensor 6 on the upper part and sensor 7 on the lower part of the casing, near the outgoing shaft). Simultaneously, an oil pump was running at a speed of 1430 RPM. The pump vane consists of 7 blades. The data was measured for 6.08 second, low-pass filtered (analog) to 5kHz and oversampled at 12.8kHz. This gives a data length of $6.08 \times 12800 = 77824$ [20].

In this paper, the analyses have been done with sensors 1 and 4 under maximal load and no load conditions. The signals have been split to 100 segments to provide 100 feature sets for each gear condition; 40% and 60% of them were used for training and testing, respectively. The remaining details are similar as section 7.1. Distribution of CWT coefficients and selected scales for the first signal segment under maximal load is shown in Fig. 14, and for an example, selected scales for first three IMFs by local Gaussian correlation are shown in Fig. 15. The final level distributions contain 42 scale levels corresponding to scales 1 through 42, shown in Fig.15 for sensor 4.

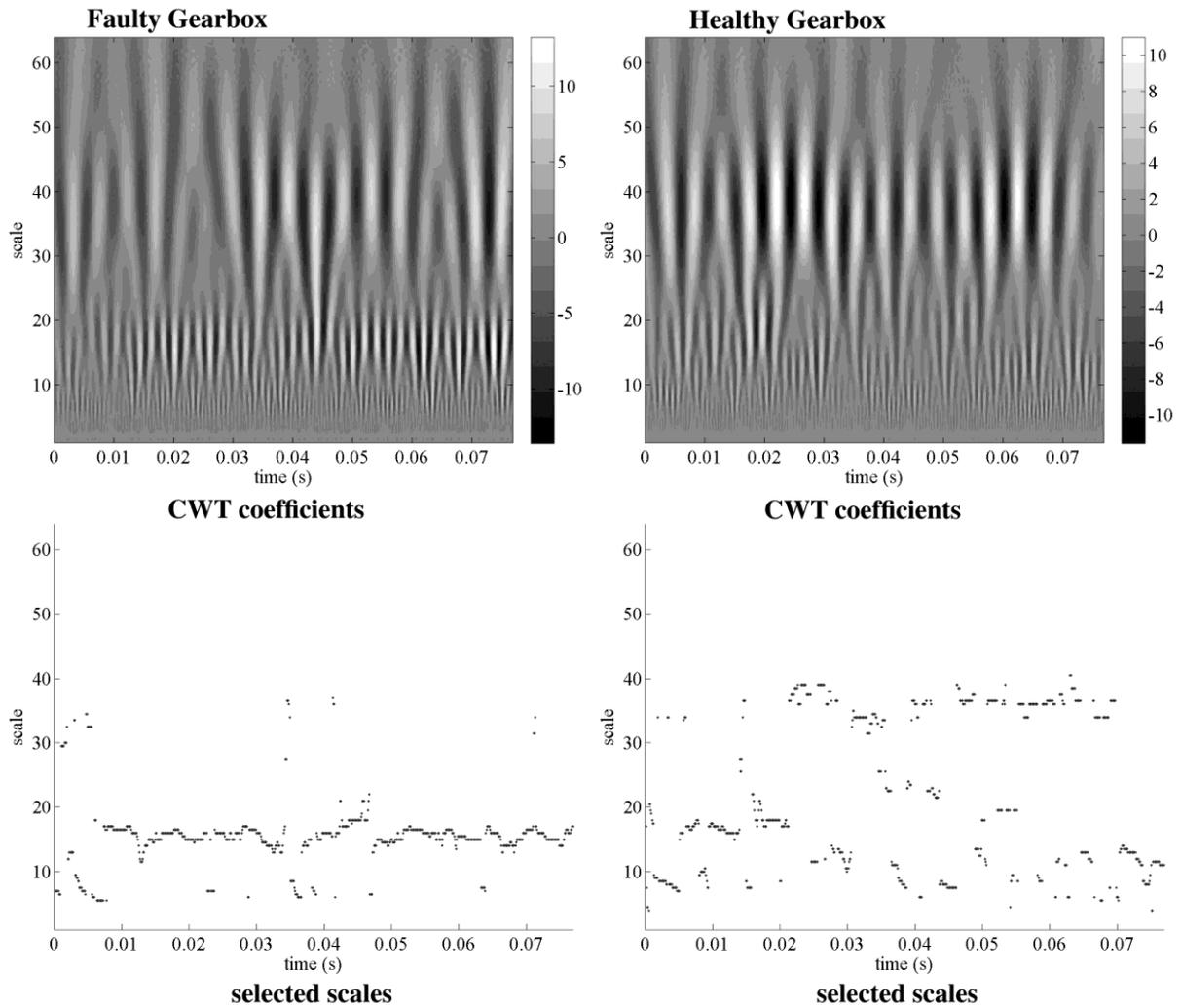

Figure 14. Distribution of CWT coefficients and selected scales
for the first signal segments with maximal load (sensor 4).

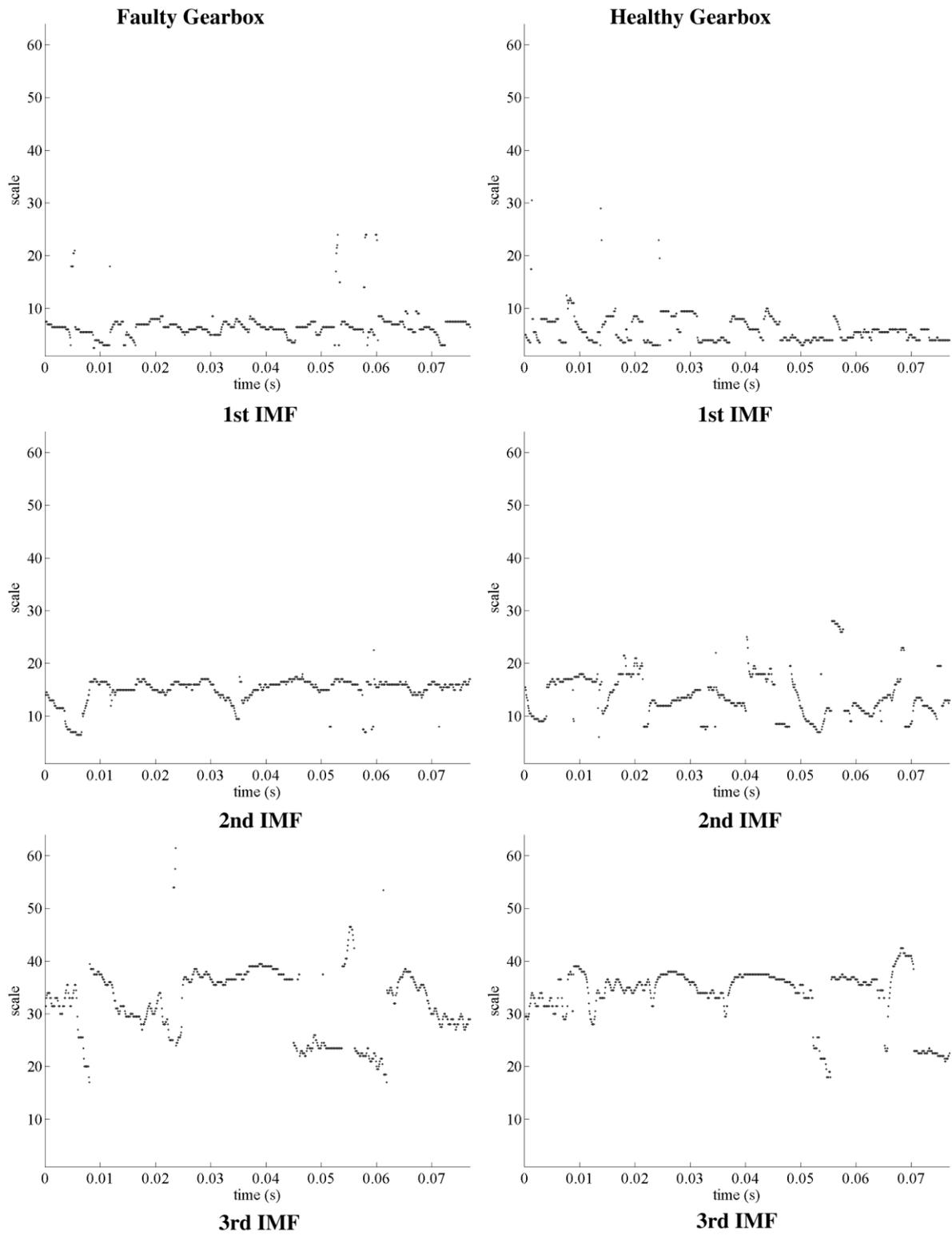

Figure 15. Selected scales by local Gaussian correlation for the first signal segments with maximal load (sensor 4).

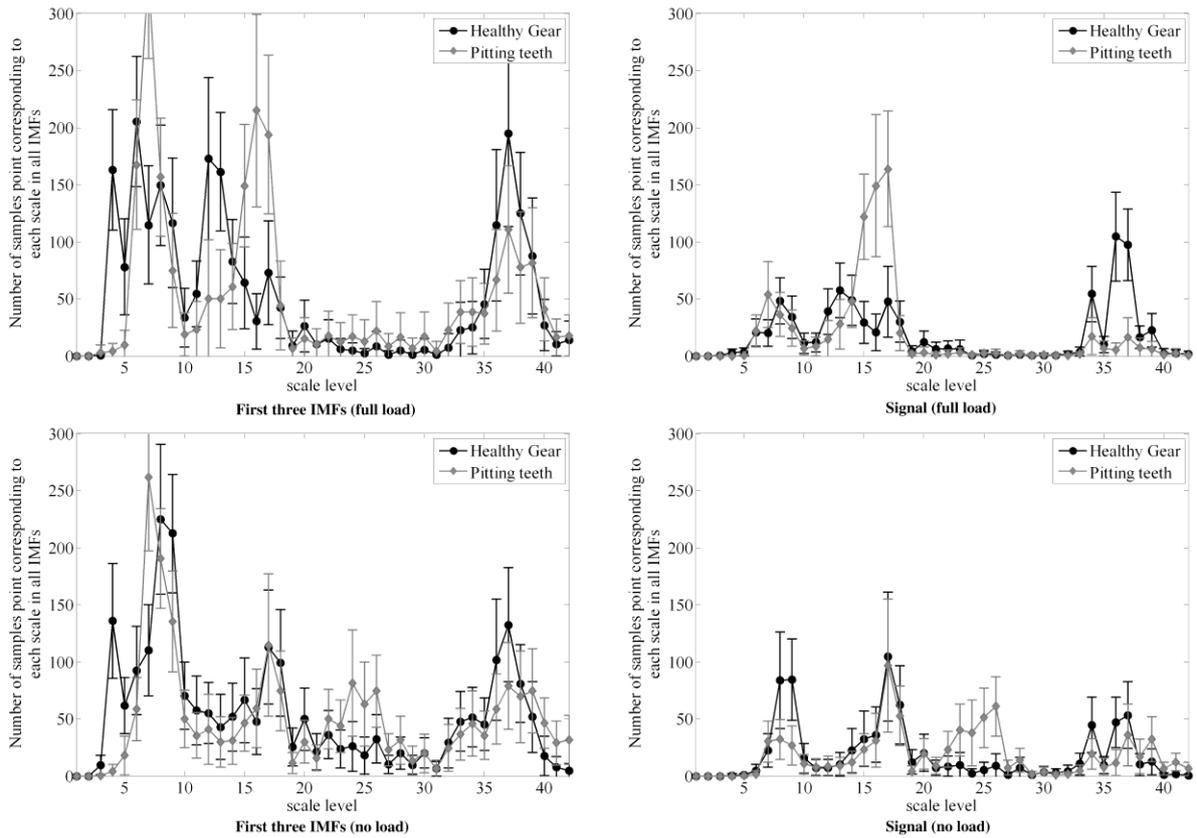

Figure 16. Mean of distribution of scale levels of all signal segments for sensor 4.

The classification performance for full and no load conditions is shown in Tables 4 and 5. As shown in these tables, by using EMD, classification performance has been increased under maximal load and no-load conditions in case of local Gaussian correlation function. The exception is on Table 5 in case of normalized dot product. It is clear from results that normalized dot product has excellent ability in feature extraction and local Gaussian correlation does not have good performance in this case. To justify the result it may be said that local Gaussian correlation and dot product focus more on the magnitude of wavelet coefficients comparing to normalized dot product and in some cases similarity between signal and wavelet coefficients maybe dominate on the magnitude of wavelet coefficients. So in these conditions, normalized dot product has better ability on feature extraction than the two other functions. With this explanation, the results of Table 4 and 5 in case of signal, the similarity of signal, and wavelet coefficient is more important than the magnitude of wavelet coefficients, this condition doesn't appear on case of decomposed signal in EMD, and local Gaussian correlation again is dominant in feature extraction.

Table 4. Classification performance of Linear-SVM (full load)

| Objective Function | Sensor Channel | First three IMF | | Signal | |
|---|---|---|---|---|---|
| | | Train success (%) | Test success (%) | Train success (%) | Test success (%) |
| Local Gaussian Correlation | 1 | 100 | 100 | 100 | 100 |
| Dot product | 1 | 100 | 100 | 100 | 100 |
| Normalized dot product | 1 | 100 | 100 | 100 | 100 |
| Local Gaussian Correlation | 4 | 100 | 100 | 100 | 97.50 |
| Dot product | 4 | 100 | 99.16 | 100 | 99.16 |
| Normalized dot product | 4 | 100 | 100 | 100 | 100 |

Table 5. Classification performance of Linear-SVM (no load)

| Objective Function | Sensor Channel | First three IMF | | Signal | |
|---|---|---|---|---|---|
| | | Train success (%) | Test success (%) | Train success (%) | Test success (%) |
| Local Gaussian Correlation | 1 | 100 | 100 | 100 | 100 |
| Dot product | 1 | 100 | 100 | 100 | 98.33 |
| Normalized dot product | 1 | 100 | 100 | 100 | 100 |
| Local Gaussian Correlation | 4 | 100 | 99.16 | 100 | 96.66 |
| Dot product | 4 | 100 | 99.16 | 100 | 99.16 |
| Normalized dot product | 4 | 100 | 98.33 | 100 | 100 |

## 8. Conclusions

In this paper, a new method of feature extraction for fault diagnosis has been presented. The method uses Gaussian correlation function in a time frame to determine appropriate scales. Gaussian correlation function can treat to non-zero mean signal in each time frame by conversion of a signal into zero-mean signal in each time frame. EMD is also presented to tackle multi-component signals in real applications. To evaluate the efficiency of the presented method in fault diagnosis of practical applications, the vibration signals, which have been acquired from two gearboxes, are used in this paper.

The obtained results indicate that, the feature extraction based on local Gaussian correlation function of wavelet coefficients and inspected signal has excellent ability to extract features and extracted features are almost linearly separable. In this study, generally local Gaussian correlation has the better performance in feature extraction; also normalized dot product has the best performance in one of the data sets without considering the IMFs. It's been said local Gaussian correlation and dot product are focused more on the magnitude of wavelet coefficients than similarity of wavelet coefficients and inspected signal comparing to normalized dot product and in signal without EMD, normalized dot product may have better performance as seen in one case.

Instead of using evolutionary optimization algorithms (such as GA, PSO), non-adaptive Morlet wavelet and discrete scales has been used in this study. It has been shown that Morlet wavelet has excellent ability to be used in exact wavelet analysis as a non-adaptive wavelet, which sped up the algorithm but did not affect extracted features.

The effects of EMD generally can improve classification performance in different applications, in some cases because of the mentioned combining features may degrade classification performance and need to be more investigated. In sense of computation time, the authors' suggestion is to prevent EMD with this procedure because the number of IMFs in processing has direct relation to computation time. In case of EMD usage, finding a procedure to consider a contribution factor for each IMF can be the subject of future investigations to enhance extracted features.

**Acknowledgements**

The Lemmer's data sets were acquired in the Delft "Machine diagnostics by neural networks"-project with help from TechnoFysica b.v., The Netherlands. The authors thank Dr. Alexander Ypma in Delft University of Technology for making the data set available for us. The authors also thank Mr. Mohammad Sarikhani, educational instructor in Amirkabir University of Technology (Vibration and Dynamics of Machine Lab.) for his valuable consultations.